\documentclass[fleqn,twoside]{article}
\usepackage{espcrc2}

\usepackage{graphicx}
\usepackage[figuresright]{rotating}

\newcommand{\PRL}{Phys.Rev.Lett.}
\newcommand{\PL} {Phys.Lett.}
\newcommand{\PR} {Phys.Rev.}

\newcommand{\AmS}{{\protect\the\textfont2
  A\kern-.1667em\lower.5ex\hbox{M}\kern-.125emS}}
\newcommand{\dE}   {\ensuremath{\mathrm{\Delta E}}}
\newcommand{\Mbc}  {\ensuremath{\mathrm{M_{bc}}}}

\newcommand{\epem} {\ensuremath{\mathrm{e^+e^-}}}
\newcommand{\Ys}   {\ensuremath{\Upsilon(\mathrm{4S})}}

\newcommand{\DSZ}  {\ensuremath{\mathrm{D^{*0}}}}

\newcommand{\DZ}   {\ensuremath{\mathrm{D^0}}}
\newcommand{\KP}   {\ensuremath{\mathrm{K^+}}}
\newcommand{\KS}   {\ensuremath{\mathrm{K^0_S}}}
\newcommand{\KM}   {\ensuremath{\mathrm{K^-}}}
\newcommand{\PM}   {\ensuremath{\mathrm{\pi^-}}}
\newcommand{\PZ}   {\ensuremath{\mathrm{\pi^0}}}

\newcommand{\Ds}   {\ensuremath{\mathrm{D_s^+}}}

\newcommand{\DsJ}  {\ensuremath{\mathrm{D_{sJ}(2317)}}}
\newcommand{\DssJ} {\ensuremath{\mathrm{D_{sJ}(2457)}}}
\newcommand{\Jpsi} {\ensuremath{\mathrm{J/\Psi}}}
\newcommand{\pipi} {\ensuremath{\mathrm{\pi^+\pi^-}}}

\newcommand{\MeV}  {\ensuremath{\mathrm{MeV}}}

\newcommand{\GeV}  {\ensuremath{\mathrm{GeV}}}

\newcommand{\fb}   {\ensuremath{\mathrm{fb^{-1}}}}
\newcommand{\nbs}  {\ensuremath{\mathrm{nb^{-1}s^{-1}}}}
\newcommand{\X}    {\ensuremath{\mathrm{X(3872)}}}
\newcommand{\K}    {\ensuremath{\mathrm{K^{\pm}}}}
\newcommand{\Ce}   {\ensuremath{\mathrm{h_c'}}}
\newcommand{\Cz}   {\ensuremath{\mathrm{\Psi_2}}}
\newcommand{\Cd}   {\ensuremath{\mathrm{\Psi_3}}}
\newcommand{\De}   {\ensuremath{\mathrm{\eta_c''}}}
\newcommand{\Dz}   {\ensuremath{\mathrm{\chi_{c1}'}}}
\newcommand{\Dd}   {\ensuremath{\mathrm{\eta_{c2}}}}
\newcommand{\cc}   {\ensuremath{c\bar{c}}}
\newcommand{\Ec}   {\ensuremath{\mathrm{\eta_c}}}
\newcommand{\Kcz}  {\ensuremath{\mathrm{\chi_{c0}}}}
\newcommand{\Eczs} {\ensuremath{\mathrm{\eta_c(2S)}}}

\title{Recent Results from Belle}

\author{Rolf Seuster\address{Department of Physics and
    Astronomy,\\
    2505 Correa Read, Honolulu, HI 96822, USA}}
       
\begin{document}

\begin{abstract}
The huge data sample accumulated at the KEKB storage ring allows for
dedicated analyses in charm spectroscopy. A new, narrow resonance with
a mass of 3.872\GeV\ was found in decays of $B$ mesons. Its properties
remain unexplained. Results on various properties of the \DsJ\
resonances have been updated. Older results on the unexpectedly large
cross section for double charmonium production in \epem\ annihilation
have been confirmed and a refined analysis is presented.
\vspace{1pc}
\end{abstract}

\maketitle
\section{Introduction}
\label{intro}
Although known as a $\mathrm{B}$-factory, the asymmetric storage ring
KEKB enables the Belle detector to explore the charm sector. Charmed
hadrons are not only the dominant decay product of $b$ flavoured
mesons, but, in addition, the decay kinematics give valuable
constraints allowing the detection of previously unobserved particles
and the determination of their quantum numbers.

Due to the excellent performance of the KEKB accelerator, the data
sample available for charm studies at Belle is unprecedented. Over a
year ago, KEKB reached its design luminosity of ${\cal
 L}=10^{-34}/\mathrm{cm^2/s}$ and since then shifted its instantaneous
peak luminosity to 13.92\nbs. Until July 2004, Belle accumulated a total
integrated luminosity of about 287~\fb. About 257~\fb\ was recorded at
the \Ys\ resonance, with the largest part of the remaining 29
\fb\ about 60 MeV\ below the resonance. In the following a few recent
results from the Belle collaboration based on smaller data sets are
presented, see \cite{BelleWeb} and references therein for more details
on the analyses.

\section{X(3872)}
In decays of $B$ mesons, the $b\to c\bar{c}s$ transition is of great
importance due to the CKM matrix ele\-ments involved. Both vertices
of the exchanged $W$ boson carry the largest possible matrix
elements for this decay. So, final states with double charm and
charmonium states can be produced to a vast amount. For charmonium,
$B$ decays turn out to be a complementary testing field to the production
of these states in \epem\ annihilation at lower energies.

In general, two variables are commonly used for selecting fully
reconstructed $B$ candidates, i.e. candidates which while decay
chain has been completely identified and reconstructed. Both make use
of the fact that at this 
center-of-mass energy (CME), both $B$ mesons are produced back-to-back
with a well known energy, $E=\sqrt{s}/2$.
One is the beam constrained mass \Mbc, where 
for the invariant mass of the $B$ candidate, the energy of the candidate is
replaced by the well known 
\mbox{energy $E$ : $\Mbc=\sqrt{E^2-\vec{p}^2_{cand}}$.}
The other one 
is the energy difference $\dE=E-E_{cand}$ between
$E$ and the energy of the $B$ candidate.

\begin{figure}[htb]
\includegraphics[width=0.425\textwidth]{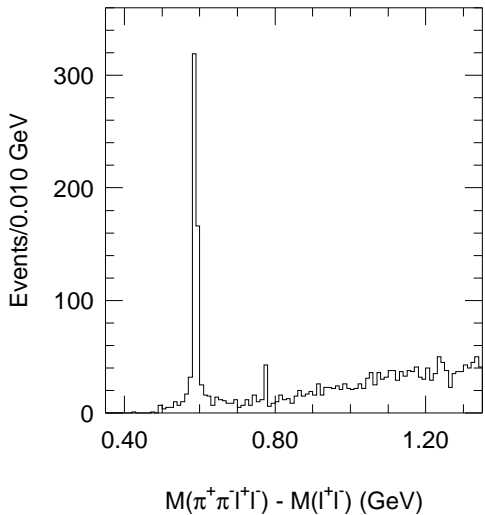}
\caption{The mass difference between the \Jpsi\pipi\ system and the
  \Jpsi\ itself. The large peak at $600~\MeV$ is due to the decay of
  the $\Psi(3770)$. The second, smaller peak at $770~\MeV$ cannot be
  explained by any known particle and has been attributed to a new
  particle, called the \X.}
\label{fig:Xdiscover}
\end{figure}

Selecting $B^\pm\to \Jpsi \pipi K^\pm$ decays by requiring \Mbc\ of the
$B$ candidate to be within $5.271~\GeV<\Mbc<5.289~\GeV$ and the energy
difference \dE\ to be within $|\dE|<30~\MeV$, a clear peak at around
$600~\MeV$ in the
mass difference between the \Jpsi\ plus the \pipi\ system and the
\Jpsi\ can be identified with the $\Psi(3770)$. A second peak
about 170~\MeV\ higher than the one from the $\Psi(3770)$ can be seen. MC
studies showed that this could not be produced by e.g. reflections of
any known particle, see Fig. \ref{fig:Xdiscover}. It has therefore
been attributed to a new particle, dubbed the X(3872), as it's
measured mass is $3872.0\pm1.0~\MeV$ \cite{BelleX}. It's resolution
has been determined to be less than the detector resolution for this
channel of $2.3~\MeV$. The product branching ratio for $B^\pm\to\X
K^\pm$ has been determined to ${\cal B}(B^\pm\to\X
K^\pm)=(1.3\pm0.3)\times10^{-5}$. Its large value suggests that the
discovery channel is one of the, if not the dominant decay channel of
the \X.

Subsequently, the \X\ was confirmed by CDF\cite{CDFX}, D0\cite{D0X}
and BaBar\cite{BaBarX}. Below, the mass and width measurements of the
\X\ for all four experiments are listed, together with its weighted
average calculated from the first three experiments only as the BaBar
result quotes only a total uncertainty. The D0 measurement 
has been transformed from $\Delta M(\pipi\Jpsi-\Jpsi)$ to $M(\X)$
using the current world average of $M(\Jpsi)$.
\begin{center}
\begin{tabular}{@{}lcc}
\hline
Exp.                & mass [\MeV] & width [\MeV] \\ \hline
Belle~\cite{BelleX} & $3872.0\pm0.6\pm0.5$ & $<2.3$ \\
CDF~\cite{CDFX}     & $3871.3\pm0.7\pm0.4$ & $<\mathrm{det.~res.}$ \\
D0~\cite{D0X}       & $3871.8\pm3.1\pm3.0$ & $<\mathrm{det.~res.}$ \\
BaBar~\cite{BaBarX} & $3873.4\pm1.4$       & - \\
average             & $3871.66\pm0.45\pm0.31$ & - \\
\hline
\end{tabular}\\[20pt]
\end{center}

\begin{figure}[htb]
\includegraphics[width=0.45\textwidth]{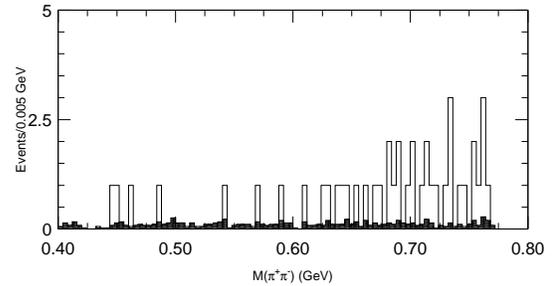}
\caption{The mass distribution of the \pipi\ system clusters at large
  values close to the kinematic limit. This suggests, it could
  originate from a $\rho^0$, which would constrain the $C$-parity of the
  new particle \X. The hatched histogram shows data from the sidebands.}
\label{fig:dipion}
\end{figure}

Various two particle invariant masses do not exhibit a noticeable
structure except the invariant mass of the \pipi\ system, see
Fig. \ref{fig:dipion}.
Similar to the corresponding distribution for the $\Psi(3770)$, the
dipion mass prefers larger values, close to the kinematic limit of
around $770~\MeV$. It's distribution suggest, the dipion system could
originate from a $\rho^0$. This would have the direct consequence of
constraining the $C$-parity of the new particle. 
Would the dipion system actually be a $\rho^0$, the X(3872) must be a
$C=+1$ state as $C=-1$ particle cannot decay into two $C=-1$ daughters.
Furthermore, in this case, the decay into $\X\to\Jpsi\pi^0\pi^0$ would
be forbidden. Results on this decay mode can be expected once more
data is accumulated.

A second noticeable fact is the absence of the decay of the \X\ into
$D\bar{D}$, a dominant decay for the considerably lighter
$\Psi(3770)$. In decays of the form $B \to D\bar{D} K$ the following limit
has been determined to be
\begin{equation}
\frac{\Gamma(\X\to D\bar{D})}{\Gamma(\X\to\pipi\Jpsi)}<7
\end{equation}
where the corresponding ratio for the $\Psi(3770)$ is
\begin{equation}
\frac{\Gamma(\Psi(3770)\to D\bar{D})}{\Gamma(\Psi(3770)\to\pipi\Jpsi)}>160.
\end{equation}
This disfavours strongly an assignment of natural quantum numbers like
$0^{++}$ or $1^{--}$ to this new state.

As the \X\ decays into a charmonium state, it is plausible to consider
it to be a charmonium as well.
Grouping all reasonable charmonium candidates, i.e. candidates with a
significantly large branching ratio into $\Jpsi\pipi$, into groups
depending on their $C$ parity leaves 3 candidates for each possible
parity assignment.
These are the \Ce, \Cz\ and \Cd\ for the $C=-1$ assignment and the \De, \Dz\
and \Dd\ for the $C=+1$ assignment. In the following, it will be shown
that the measured properties of the \X\ disfavour all six possible
charmonium assignments listed above.
As for the \Cd, a large branching fraction into $DD^*$ is expected.
However, it has been argued, that the large angular momentum strongly
suppresses this decay mode.
\begin{tabular}{@{}llll}
\hline
$C$-parity & Name & $J^{PC}$ & spectroscopic name\\ \hline
     & \Ce  & $1^{+-}$ & $2^1P_1$ \\
C=-1 & \Cz  & $2^{--}$ & $1^3D_2$ \\
     & \Cd  & $3^{--}$ & $1^3D_3$ \\ \hline
     & \De  & $0^{-+}$ & $3^1S_0$ \\
C=+1 & \Dz  & $1^{++}$ & $2^3P_1$ \\
     & \Dd  & $2^{-+}$ & $1^1D_2$ \\
\hline
\end{tabular}\\[20pt]

\subsection{Is $C(\X)=-1$ ?}
One obvious candidate for the \X\ is the $h_c'$, for which a
sizable branching ratio into \pipi\Jpsi\ is expected and its predicted
mass is rather close to the mass of the \X. Here, a distribution
worthwhile to check is the angular distribution of the \K\ in the rest
frame of the \X\ with respect to the flight direction of the \Jpsi\ in
the rest frame of the \X\ candidate. For the $h_c'$ as well as any
other $1^{+-}$ state, the distribution will be proportional to
$\sin^2\theta$. However, the measured distribution shows clearly a different
behaviour as indicated by the reduced $\chi^2$ of $\chi^2/d.o.f.=75/6$.

As for the \Cz\ and \Cd, again, the measured \pipi\ distribution does not
agree with the expected distribution for these two particles.
Additionally, for these particles, radiative decays into $\gamma
\chi_{c1}$ and $\gamma \chi_{c2}$ would have
branching ratios more than two and a half times larger or more than
three and a half times larger than that of the discovery mode,
respectively. However, no signal has been found and upper limits on the ratio
of branching ratios have been determined to be
\begin{equation}
\frac{\X\to \gamma\chi_{c1}}{\X\to\pipi\Jpsi}<0.9
\end{equation}
and
\begin{equation}
\frac{\X\to \gamma\chi_{c2}}{\X\to\pipi\Jpsi}<1.1
\end{equation}
respectively. This rules out also these two assignments.

In conclusion, no good charmonium candidate with $C(\X)=-1$ is left.
\subsection{Is C(\X)=+1 ?}
This would mean that the \pipi\ system is a $\rho^0$. However, the decay
would be parity-violating and therefore strongly suppressed.
Why the discovery channel has such a large branching ratio, would
still have to be explained.

Here, the candidates are the \De, \Dz\ and \Dd. 

The mass of the \De\ is expected to be in the order of $\sim
4000~\MeV$, as it should be about approximately $40~\MeV$ below its
partner, the $\Psi(3S)$ with $m(\Psi(3S))=(4040\pm2)~\MeV$.

For the \Dz, the radiative decay into $\gamma \Jpsi$ is expected to
dominate over the discovery channel. No signal is observed, strongly
disfavouring this possible assignment.

For the \Dd, it is expected that the decay into $\Dd\to
\eta_c\pipi$ dominates over the discovery channel. However, the large
branching ratio of the discovery channel would result in a extremely
large branching ratio for this decay mode. This makes this assignment
also very unlikely. 

\subsection{New Type of Matter?}
One striking fact however remains: The mass of the \X\ is within
the small uncertainties identical to the sum of the masses of the \DZ\
and the \DSZ. Many years ago such molecule-type bound states of mesons
were predicted, see e.g. references in \cite{BelleX}.

With more data, the nature of this new particle will be more explored.

\section{\DsJ\ and \DssJ\ }
Over a year ago, more particles were discovered, which didn't fit well
into the existing quark potential models. In \epem\ annihilation,
BaBar \cite{BaBarDsJ} claimed the discovery of a narrow state with a mass
of $2317~\MeV$ decaying into $\Ds\PZ$ and an unexplained excess in
$\Ds\PZ\gamma$ around a mass of $2460~\MeV$. Later, CLEO 
\cite{CLEODsJ} confirmed the first peak and established the second as
another new particle decaying into $\DssJ\to\Ds^*\pi^0$,
$\Ds^*\to\Ds\gamma$. Belle then confirmed both particles and
identified them also in decays of $B$ mesons \cite{BelleDsJ,BelleDsJB}.

According to existing quark potential models, the masses of these two
particles should be more than $50~\MeV$ higher than the missing meson
states containing $c\bar{s}$, the ${}^3P_0$ and the admixture of the
${}^1P_1$ and the ${}^3P_1$.
The current world averages of these two new states are 
$(2317.4\pm0.9)~\MeV$ and $(2459.3\pm1.3)~\MeV$, respectively.

Except for the too low masses, the other properties are in a good
agreement with the expectation for these two $c\bar{s}$ candidates.

As both particles are too light to decay into $\mathrm{DK}$ and
$\mathrm{D^*K}$, respectively, their dominant decay, assuming they are
the missing $c\bar{s}$ states, will be the modes in which they have
been discovered. Furthermore, the width of radiative and other decays
are with the current, limited statistics, well in agreement with the
theoretical expectations of Bardeen et al. \cite{Bardeen} and by Godfrey
\cite{Godfrey}. Only one upper limit by CLEO on the ratio of
$\frac{DsJ\to\Ds^*\gamma}{DsJ\to\Ds\pi^0}<0.06$
cannot be explained by above mentioned models. The only other upper
limit on this ratio by Belle determined in decays of $B$ mesons, is
significantly weaker. Except for the discovery channels, only two
other decay modes have been observed, both for the \DssJ. These are
$\DssJ\to\Ds\gamma$ seen by Belle in both $B$ decays $(0.38\pm0.13)$
and in \epem\ annihilation $(0.55\pm0.15)$ as well as by BaBar in $B$
decays only $(0.38\pm0.12)$. Latter one is the weighted average over
various $B$ decay modes.
The other observed decay mode of the \DssJ\ is the decay
$\DssJ\to\Ds\pipi$ with an relative branching ratio of $(0.14\pm0.05)$
w.r.t. to discovery channel, observed in \epem\ annihilation by
Belle. This number slightly, i.e. with a significance of
about $1~\sigma$, disagrees with the two upper limit set by Belle in
$B$ decays $(<0.10)$ and by CLEO $(<0.08)$.
Other upper limits on decay modes are also in the (10-60)\% order, and
agree rather well with the predictions.

For the lighter \DsJ, only predictions for the decay $\DsJ\to\Ds^*\gamma$
exist. Here, the prediction by Bardeen is closer to the stringent
limit by CLEO mentioned above, the prediction by Godfrey is higher by
a factor of three. 
Experimentally, two more upper limits for decays of the \DsJ\ have
been determined, however, no theory predictions for these two modes
exist. These two decay modes are $\DsJ\to\Ds\gamma$ and
$\DsJ\to\Ds\pipi$, which are forbidden for the missing $c\bar{s}$
states. Stringent limits in the percent or even the permille level
support the hypothesis, this particle is in fact the so far
unobserved $c\bar{s}$ state.

More information about the spin of an particle can be derived from
angular distributions like helicity distributions, decay angle
distributions in the rest frame of the mother particle w.r.t. the
flight direction of the mother in the rest frame of the $B$.


For the \DsJ\ the helicity angle distribution for the decay
$\DsJ\to\Ds\pi^0$ is in excellent agreement with the spin 0
hypothesis
.

Also for the \DssJ, the helicity angle distribution in
$\DssJ\to\Ds\gamma$ supports the spin 1 assignment of this particle,
likely being the so far missing $c\bar{s}$ state
.

In conclusion, the new discovered particle, named the \DsJ\ and the
\DssJ, behave like the unobserved $c\bar{s}$ states, however their
masses are significantly too low. 

\section{Double $c\bar{c}$ production}
Another very nice and elegant analysis showing surprising results is
the investigation about the double \cc\ production. Here, a \Jpsi\ is
reconstructed in the very clean lepton channel, then, simply, its
recoil mass is plotted. The effects of ISR photons are taken into
account, see \cite{BelleCC1,BelleCC2} for details.
A momentum cut removes contribution from decays of $B$ mesons.

\begin{figure}[htb]
\vspace{9pt}
\includegraphics[width=0.55\textwidth]{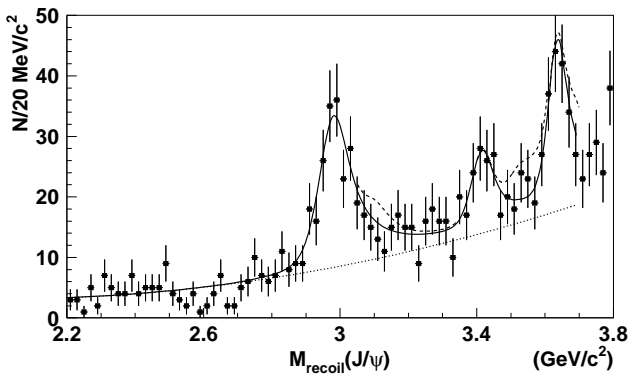}
\caption{The recoil mass spectrum against a fully reconstructed \Jpsi\
in \epem\ annihilation events. A clear enhancement of activity above
the $c\bar{c}$ threshold is visible, in contrast to predictions by
NRQCD.}
\label{fig:recoil}
\end{figure}

The recoil mass spectrum does not reveal a large signal until about
$3~\GeV$, the charm production threshold, where the signal strongly
increases.

Zooming into the threshold region, see Fig. \ref{fig:recoil}, reveals
three clear distinct peaks, 
in a previous analysis \cite{BelleCC1} identified with the \Ec,
\Kcz\ and the \Eczs. The masses of these particles were consistent
with the world average. For the \Eczs, it was only the second
observation.
Note, that e.g. the production of two \Jpsi\ from a single $\gamma$
is forbidden by conservation of $C$-parity.

Another surprising result of the old analyses were that there is
almost no activity below $3~\GeV$, but a very strong activity above.

So far, the only theory predicting cross sections of charmonium
production at this center-of-mass energy is non-relativistic QCD,
NRQCD for short. It predicts the $\Jpsi gg$ production to be dominant
with a total cross section of about $\sim1~\mathrm{pb}$. The colour
singlet production of $\Jpsi g$ might dominate the phase space as a
quasi two body decay at the endpoint with up to
$\sim0.5~\mathrm{pb}$. Double \cc\ production via gluon splitting into
\cc\ is expected to be suppressed by about a factor of 10-20 w.r.t. to
$\Jpsi gg$.

The measurement however finds the $\Jpsi gg$ process to be negligible
and the double \cc\ production dominates with about 80\%. The total
cross section for double \cc\ production is about one order of
magnitude too large compared to the prediction of NRQCD.

Several, alternative explanations arose, e.g. that under the \Ec\
peak, significant contribution from
$\epem\to\gamma\gamma\to\Jpsi\Jpsi$ could be hidden. Glueballs could
interfere with the charmonium states and increase the cross section,
or, a bias in the momentum scale can deteriorate any interpretation,
that the first peak would no be due to the \Ec.

In a refined analysis \cite{BelleCC3} performed on a larger data set
of $155~\fb$, the fit to the recoil mas spectrum now allowed for all
narrow resonances in this mass region. To the \Ec, $\chi_{c0}$, \Eczs\
allowed in the previous analysis, the \Jpsi, $\chi_{c1(2)}$ and the
$\Psi(2S)$ were added. However, the fitted yield for the last three particles
was consistent with zero, even negative for the \Jpsi\ and the
$\Psi(2S)$. As no signal was observed, the mass and width of these
particles were fixed to their world average values. For the other
three particles, these values were floated in the fit. The fitted
yields are listed below

\begin{tabular}{@{}lcrc} \hline
$c\bar{c}$     & mass [\MeV] & N          & signif. \\ \hline
\Ec            & $2972\pm7$  & $235\pm26$ & 10.7 \\
\Jpsi          & fixed       & $-14\pm20$ & - \\
$\chi_{c0}$    & $3407\pm11$ & $89\pm24$  & 3.8 \\
$\chi_{c1(2)}$ & fixed       & $10\pm27$  & - \\
\Eczs          & $3630\pm8$  & $164\pm23$ & 5.3 \\
$\Psi(2S)$     & fixed       & $-26\pm29$ & - \\ \hline
\end{tabular}
\\
In order to confirm the peak around $3~\GeV$ is caused by \Ec\ and
not by \Jpsi, various tests have been performed. First, about
230 signal events in the \Ec\ mass region allow fully
reconstruct a few events. The \Ec was reconstructed in the
$\Ec\to\KS\KP\PM$ and the $\Ec\to2(\KP\KM)$ 
decays. From Monte-Carlo, 2.6 events were expected to be
reconstructed. In the data, 3 events have been observed, in excellent
agreement with the expectation. Additionly, in the dilepton channel of
the \Jpsi\ no event events has been observed, in accordance with the
expectation.

Furthermore, an angular analysis of the production and helicity angle
distribution for the reconstructed \Jpsi\ have been performed for all
events in the three signal peaks. For the production via
$\epem\to\gamma^*\to\Jpsi\cc$, due to conservation of angular
momentum of the exchanged $\gamma^*$, the production and the helicity
angle distributions have to be equal. For processes like
$\epem\to\gamma\gamma\to\Jpsi\cc$ no such constrain exist.

All distributions have been fitted with a function of the form
$f(x)\sim1+\alpha\cos^2{\theta}$ with $\alpha$ floated. Within large
uncertainties, the results are consistent with the hypothesis of a
$1~\gamma$ exchange. The last column in the table below actually
assumes the production via the $1~\gamma$ exchange, reducing slightly
the uncertainty on $\alpha$. Also listed in the last row are the
predictions for a glueball.

\begin{tabular}{@{}lrrr} \hline
          & $\alpha_{prod}$
          & $\alpha_{hel}$
          & $\alpha_{prod}=\alpha_{hel}$ \\ \hline
\Ec       & $1.4^{+1.1}_{-0.8}$
          & $0.5^{+0.7}_{-0.5}$
          & $0.93^{+0.57}_{-0.47}$ \\
$\chi_{c0}$&$-1.7^{+0.5}_{-0.5}$
          & $-0.7^{+0.7}_{-0.5}$
          & $-1.01^{+0.38}_{-0.33}$ \\
\Eczs     & $1.9^{+2.0}_{-1.2}$
          & $0.3^{+1.0}_{-0.7}$
          & $0.87^{+0.86}_{-0.63}$ \\
G($0^+$)  & -0.9 & -0.9 \\ \hline
\end{tabular}
\\
Additionly, the cross sections for the production of double charmonium
have been determined. The results
are listed in the table below. Note, that for the decay of the charmonium
accompanying the \Jpsi, two or more charged particles were required
in the analysis, reducing the visible cross section and therefore
enhancing even more the discrepancy between theory prediction and
experiment.

\begin{tabular}{@{}lccc} \hline
$c\bar{c}$     & $\sigma_{Born}\times {\cal B}(c\bar{c}\to \ge2\mathrm{charged})[fb]$ \\ \hline
\Ec            & $25.6 \pm 2.8 \pm 3.6$ \\
\Jpsi          & $< 9.1$ \\
$\chi_{c0}$    & $6.4 \pm 1.7 \pm 1.0$ \\
$\chi_{c1(2)}$ & $< 5.3$ \\
\Eczs          & $16.5\pm3.0\pm2.5$ \\
$\Psi(2S)$     & $< 13.3$ \\ \hline
\end{tabular}
\\

\section{Conclusion}
Over the last years, the $b$-factories made several new unexpected
discoveries, mostly involving charmed mesons. An apparent
charmonium state, the \X, does not fit at all into the charmonium
spectrum. It might represent a new form of matter, a molecule build
out of mesons. Other new discovered particles in the $c\bar{s}$ system
behave like expected, however, their masses are significantly too
low. And last, but not least, the production cross section of double
charmonium in \epem\ annihilation is still about an order of magnitude
larger than predicted by theory.



\begin{thebibliography}{9}
\bibitem{BelleWeb}
  http://belle.kek.jp/bdocs/b\_journal.html
\bibitem{BelleX} S.-K.~Choi, S.L.~Olsen, et {\it al.}(Belle Coll.)
                 \PRL {\bf 91} (2003) 262001.
\bibitem{CDFX}   hep-ex/0312021
\bibitem{D0X}    hep-ex/0405004
\bibitem{BaBarX} hep-ex/0406022

\bibitem{BaBarDsJ}  B.~Aubert, et {\it al.}(BaBar Coll.)
                    \PRL {\bf 90} (2003) 242001.
\bibitem{CLEODsJ}   D.~Besson, et {\it al.}(CLEO Coll.)
                    \PR  {\bf D68}(2003) 032002.
\bibitem{BelleDsJ}  Y.~Mikami, et {\it al.}(Belle Coll.)
                    \PRL {\bf 92} (2003) 012002.
\bibitem{BelleDsJB} P.~Krokovny, A.~Bondar, et {\it al.}(Belle Coll.)
                    \PRL {\bf 91} (2003) 262002.
\bibitem{Bardeen}   W.A.~Bardeen, et {\it al.}
                    \PR  {\bf D68}(2003) 054024.
\bibitem{Godfrey}   S.~Godfrey, et {\it al.}
                    \PL  {\bf B568}(2003) 254.
\bibitem{BelleCC1}  K.~Abe, et {\it al.} (Belle Coll.)
                    \PRL {\bf 89} (2002) 142001 
\bibitem{BelleCC2}  K.~Abe, et {\it al.} (Belle Coll.)
                    \PR {\bf D70} (2004) 071102(R).
                    
\end{thebibliography}
\end{document}